\documentclass[aps,prl,twocolumn,amsfonts,amssymb,amsmath,10pt,showpacs,showkeys]{revtex4-1}

\usepackage{graphicx}%
\usepackage{dcolumn}%
\usepackage{bm}%
\usepackage{bbm} 
\usepackage[usenames]{color}
\usepackage{soul} % highlights text
\usepackage[normalem]{ulem} %usage \sout{...}
\usepackage{endnotes}
\usepackage{color}
\usepackage{ulem}

\begin{document}

\title{Investigation of Feshbach Resonances in ultra-cold \textsuperscript{40}K spin mixtures}

\author{J.~S.~Krauser$^{1,2}$}
\author{J.~Heinze$^{1,2}$}
\author{S.~G\"otze$^{1}$}
\author{M.~Langbecker$^{1,3}$}
\author{N.~Fl\"aschner$^{1}$}
\author{L.~Cook$^{4}$}
\author{T.~M.~Hanna$^{5}$}
\author{E.~Tiesinga$^{5}$}
\author{K.~Sengstock$^{1,2}$}
\author{C.~Becker$^{1,2}$}
\email[Corresponding author: ]{cbecker@physnet.uni-hamburg.de}

\affiliation{
     			$^{1}$Institut f\"ur Laser-Physik, Universit\"at Hamburg, Luruper Chaussee 149, 22761 Hamburg, Germany \\
			$^{2}$Zentrum f\"ur Optische Quantentechnologien, Universit\"at Hamburg, Luruper Chaussee 149, 22761 Hamburg, Germany \\
			$^{3}$Institut f\"ur Physik, Johannes Gutenberg Universit\"at Mainz, Staudingerweg 7, 55128 Mainz, Germany \\
			$^{4}$Department of Physics and Astronomy, University College London, Gower Street, London, WC1E 6BT, United Kingdom \\
			$^{5}$Joint Quantum Institute, National Institute of Standards and Technology and the University of Maryland, Gaithersburg, Maryland 20899, USA}

%\pacs{67.85.Lm; }
%\keywords{Quantum gases; Degenerate Fermi gases; Spin waves; High spin systems; Fundamental excitations;}

\begin{abstract}
Magnetically-tunable Feshbach resonances are an indispensable tool for
experiments with atomic quantum gases.
We report on twenty thus far unpublished Feshbach resonances and twenty one further probable Feshbach resonances in spin mixtures of ultracold fermionic
$^{40}$K with temperatures well below 100 nK. 
In particular, we locate a broad resonance at $B=389.6\,\text{G}$ with a magnetic width of $26.4\,\text{G}$. Here $1\,\text{G}=10^{-4}\,\text{T}$. 
Furthermore, by exciting low-energy spin waves, we demonstrate a novel means to precisely determine the zero crossing of the scattering length for this broad Feshbach resonance. 
Our findings allow for further tunability in experiments with ultracold $^{40}$K quantum gases. 
\end{abstract}

\maketitle

Ultracold fermionic atomic gases are ideally suited for the study
of many-body quantum phenomena owing to the unrivaled control over
experimental parameters such as the spatial geometry of confining
potentials and the interaction strength between the atoms.  The
interaction strength is controlled using magnetically-tunable Feshbach
resonance and typically characterized by the $s$-wave scattering length,
which can be set to a wide range of values, either negative or positive.
Feshbach resonances have been found in many bosonic as well as fermionic
atomic systems (see \cite{Timmermans1999, Duine2004, Koehler2006,
Chin2010, Frisch2014} and references therein). The isotope $^{40}$K
constitutes one of the work horses in current experiments with
ultracold fermions and provides a rich ground-state structure 
allowing for the realization of binary and multi-component spin mixtures \cite{Regal2004, Chin2004, Partridge2005, Zwierlein2005, Joerdens2008, Schneider2008, Ospelkaus2010, Jo2009, Zhang2011, Conduit2011, Pekker2011, Nascimbene2010, Sommer2011, Koschorreck2013, Krauser2012, Heinze2013, Krauser2014, Ebling2014}
as well as several Bose-Fermi \cite{Roati2002, Goldwin2004, Ospelkaus2006a, Ospelkaus2006b, Gunter2006, Best2009, Park2012}
and Fermi-Fermi mixtures \cite{Taglieber2008, Wille2008, Gierke2010, Costa2010, Ridinger2011}. 
In the energetically-lowest hyperfine manifold with 
total angular momentum $f=9/2$ ten magnetic spin states are available
ranging from $m=-9/2,\ldots,+9/2$ and 45 binary spin mixtures
can be realized \cite{footnote1}.  So far, only three Feshbach resonances
have been reported, one for each of collision channels $\{m_1,m_2\}=
\{-9/2,-7/2\}$ \cite{Loftus2002}, $\{-7/2,-7/2\}$ \cite{Regal2003a,
Ticknor2004} and $\{-9/2,-5/2\}$ \cite{Regal2003b}.

Here, we report on the experimental observation of 20 theoretically confirmed and 22 further probable magnetic Feshbach resonances in different spin mixtures of ultracold $^{40}$K.
Their positions are determined from the enhanced, resonant loss of atoms near
the resonance.  In addition, we introduce a novel method for precisely
determining the sign changes of the scattering length around the Feshbach
resonance by exciting low-energy spin waves.
In particular, this approach enables us to measure the zero crossing with high accuracy.
We also find that the measured positions of the assigned Feshbach resonances agree well with theoretical
calculations based on multi-channel quantum defect theory using the best
available Born-Oppenheimer potentials for $^{40}$K \cite{Falke2008}.

A Feshbach resonance occurs when two atoms in well-defined spin states
collide and couple to a virtual molecular state with a different spin
configuration \cite{Timmermans1999, Duine2004, Koehler2006, Chin2010,
Frisch2014}. 
As these configurations have different magnetic moments
their relative Zeeman energy can be tuned with a magnetic field.
This leads to a magnetic-field-dependent complex scattering length
$\tilde{a}(B)=a(B)-ib(B)$, where real-valued $a(B)$ and $b(B)>0$ describe elastic and inelastic
two-body processes, respectively. 
Here, we have allowed for inelastic transitions to spin configurations whose Zeeman energy is below
that of the entrance configuration. 
In fact, near a resonance and in the limit of zero collision energy
\begin{equation} 
a(B) = a_{\text{bg}} \left( 1 - \frac{\Delta B(B-B_\text{res})}
             {(B-B_\text{res})^2+(\gamma_2/2)^2} \right) \,,
\label{eq:feshbachresonance}
\end{equation}
with resonance position $B_\text{res}$, magnetic width $\Delta
B$, and  background scattering length $a_{\text{bg}}$.
Finally $\gamma_2$ describes two-body decay to other spin-channels
(expressed in units of the magnetic field). 
Similarly, we have
\begin{equation} 
b(B)=2a_{\text{res}} \,\frac{(\gamma_2/2)^2}{(B-B_\text{res})^2+(\gamma_2/2)^2} \,,
\label{eq:feshbachresonancedamp}
\end{equation} 
with resonance length $a_{\text{res}}\,=\,a_{\text{bg}}\Delta B / \gamma_2$. 
Atom loss, quantified by the two-body rate coefficient $K_2(B)= 4\pi\hbar\,
b(B)/\mu$, is largest in the vicinity of the resonance position $B_\text{res}$. 
Here, $\hbar$ is the reduced Planck constant, $\mu=m/2$, and $m$ is the atomic mass.

We begin our experiments by preparing a spin mixture of $m_1=+9/2$ and
$m_2=+7/2$ atoms with about $N=5{\times}10^4$ atoms per spin state
in an optical dipole trap.  
The trap is harmonic and nearly isotropic with mean trapping frequency $\bar{\omega}= 2\pi\times50\,\text{Hz}$
and temperature $T\approx 0.3\,T_\text{F}$, where
$T_\text{F}=\hbar\bar\omega(6N)^{1/3}/k\approx 170\,\text{nK}$ is the Fermi temperature and $k$
is the Boltzmann constant. 
For the investigation of different collision channels, the corresponding spin mixture is prepared at a magnetic field of $B\,{=}\,45\,\text{G}$ using radio-frequency sweep protocols optimized for each mixture.
After this preparation, the magnetic field is ramped to its final value and the ensemble is held
for a time of $100\,\text{ms}$.
The magnetic field is calibrated by radio frequency spectroscopy resulting in an uncertainty of $\Delta B \le 0.2\,\text{G}$.
Subsequently, the magnetic field is switched off and the remaining atoms are counted after a time-of-flight in a Stern-Gerlach gradient field.
The field value where atom loss is maximal, $B_\text{loss}$, is assigned as the resonance position $B_\text{res}$. 
The full-width-half-maximum magnetic width of the experimental loss feature is denoted by $\sigma_{\rm loss}$.

We have located 41 resonant loss features in
this manner.  Based on multi-channel quantum defect theory
\cite{Chin2010,Hanna2009} we find that nineteen of these features are Feshbach resonances with $s$ partial-wave character and
one has $p$-wave character.  In addition, there is qualitative agreement
between the location of these experimental loss maxima and the theoretical
resonance location.  Table~\ref{feshbachtab} lists these twenty assigned
resonances.  The remaining loss features are probable Feshbach resonances.
They are listed in Table~\ref{losstab}.  For these features we have no
corresponding theoretical calculations.  Parallel to this work, groups
in Munich and Amsterdam have measured other $^{40}$K Feshbach resonances
in different collision channels \cite{privatecommunication}. 

\begin{table}[t]
\centering{
\begin{tabular}{c|c|c|cc|cc}
\toprule
$\{m_1,m_2\}$					& $M$  					& pw 	&$B_\text{loss}		$	&	$\sigma_\text{loss}$		&	$B_\text{res}^\text{calc}$		& $\Delta B^\text{calc}$	\\
\toprule
$	\{{+}1/2,{-}1/2\}		$		&$\vphantom{+}0		$	&	s 	&$	15\,(4)		$	&	$	4		$		&	$	17.3	$				&	$	0.1	$\\	
$					$		&$\vphantom{+}0		$	&	s 	&$	31\,(4)		$	&	$	5		$		&	$	30.9	$				&	$	0.2	$\\	
$					$		&$\vphantom{+}0		$	&	s 	&$	53\,(4)		$	&	$	5		$		&	$	53.4	$				&	$	0.4	$\\	
$					$		&$\vphantom{+}0		$	&	s 	&$	88\,(4)		$	&	$	4		$		&	$	87.1	$				&	$	0.4	$\\ 
$					$		&$\vphantom{+}0		$	&	s 	&$	246\,(0.8)		$	&	$	2.4		$		&	$	246.6$				&	$	2.0	$\\	
$					$		&$\vphantom{+}0		$	&	s 	&$	389\,(1)		$ 	&	$	5.5		$		&	$	389.6$				&	$	26.4	$\\	
$	\{{+}3/2,{-}3/2\}		$		&$\vphantom{+}0		$	&	s 	&$	95\,(4)		$	&	$	23		$		&	$	93.8	$				&	$	1.8	$\\	
$					$		&$\vphantom{+}0		$	&	s 	&$	182\,(4)		$	&	$	12		$		&	$	181.5$				&	$	2.2	$\\	
$	\{{+}5/2,{-}5/2\}		$		&$\vphantom{+}0		$	&	s 	&$	61\,(4)		$	&	$	21		$		&	$	61.5	$				&	$	4.2	$\\	
$	\{{+}7/2,{-}7/2\}		$		&$\vphantom{+}0		$	&	s 	&$	34.3\,(0.8)		$	&	$	10.8		$		&	$	35.0	$				&	$	3.4	$\\	
$					$		&$\vphantom{+}0		$	&	s 	&$	147.1\,(3.0)	$	&	$ 	0.8		$		&	$	145.8$				&	$	0.2	$\\	
$	\{{+}9/2,{-}9/2\}		$ 		&$\vphantom{+}0		$	&	s 	&$	17.6\,(0.3)		$	&	$	5.4 		$		&	$	18.8	$				&	$	1.8	$\\	
$	\{{+}9/2,{-}7/2\}		$		&${+}1				$	&	s 	&$	13.9\,(0.2)		$	&	$	1.3 		$		&	$	14.5	$				&	$	0.4  	$\\	
$					$		&${+}1				$	&	s 	&$	28.4\,(0.3)		$	&	$	6.1 		$		&	$	30.2	$				&	$	2.9   $\\	
$					$		&${+}1				$	& 	p 	&$	139\,(1)		$	&	$	20		$		&	$	138	$				&	$		$\\	
$	\{{+}9/2,{-}5/2\}		$		&${+}2				$	&	s 	&$	27.3\,(0.3)		$	&	$	4.8	 	$		&	$	27.1	$				&	$ 	1.4    $\\	
$					$		&${+}2				$	&	s 	&$	63.4\,(0.7)		$	&	$	30		$		&	$	63.3	$				&	$	6.0    $\\	
$	\{{+}9/2,{-}3/2\}		$		&${+}3				$	&	s 	&$	53\,(4)		$	&	$	14 		$		&	$	52.4	$				&	$	3.1    $\\	
$					$		&${+}3				$	&	s 	&$	137\,(8)		$	&	$	53		$		&	$	140.5$				&	$	14.0	 $\\	
$	\{{+}9/2,{-}1/2\}		$		&${+}4				$	&	s 	&$	114\,(8)		$	&	$	{>}40	$		&	$	113.5$				&	$	7.8    $\\	
\end{tabular}
}
\caption{
\textbf{Twenty assigned Feshbach resonances in \textsuperscript{40}K.}
The first five columns represent the collision channel $\{m_1,m_2\}$
in the $f=9/2$ manifold, the total spin quantum number $M= m_1+m_2$, the
partial wave (pw), the experimental maximum loss position $B_\text{loss}$,
and the observed magnetic width $\sigma_\text{loss}$, respectively.
The uncertainty of $B_\text{loss}$, given in parenthesis, is a one standard
deviation uncertainty with combined systematic and statistical error.  The last two
columns show the calculated resonance position $B_\text{res}^\text{calc}$
and the magnetic width $\Delta B^\text{calc}$.  
Magnetic fields and widths are in Gauss.
}
\label{feshbachtab}
\end{table}

Note that the position of a resonance determined
from atom loss measurements contains systematic deviations as reported previously, i.e. $B_\text{res}\neq B_\text{loss}$ \cite{Zhang2011}.
Similarly $\sigma_{\text{loss}}$ does not coincide with the calculated $\gamma_2$.
Atom loss is not only due to two-body collisions but is also caused by three-body
recombination, where three colliding atoms react to produce a hot molecule.
The field-dependent recombination rate coefficient $K_3(B)$, does not need to peak at
the same $B$-field or have the same width as $K_2(B)$. 
In addition, for quantum degenerate Fermi gases of $^{40}$K atoms, collective phenomena
can modify the resonance feature especially when the scattering length $a(B)$ is
large compared to $1/k_{\rm F}$, where the Fermi wavevector $k_{\rm F}$ is defined by
$\hbar^2k^2_{\rm F}/(2m)=kT_{\rm F}$ \cite{Trotzky2015}.
Finally, lineshapes can also be distorted when a large fraction of the atoms is lost.

\begin{table}[t]
\centering{
\begin{tabular}{c|c|c|cc}
\toprule
$\{m_1,m_2\}$					& $M	$				&pw			 &$B_\text{loss}$	&	$\sigma_\text{loss}$		\\
\toprule

$	\{{+}5/2,{-}9/2\}		$		&${-}2			$	&s			&$	24.0(0.5)  	$	&	$	4.2	$	\\	
$	\{{-}1/2,{-}1/2\}		$		&${-}1			$	&p			&$	373(2)	$	&	$	2	$	\\
$	\{{+}1/2,{-}1/2\}		$		&$\vphantom{+}0	$	&s			&$	61(4)		$	&	$	4	$	\\	
$	\{{+}3/2,{-}3/2\}		$		&$\vphantom{+}0	$	&s			&$	31(4)		$	&	$	6	$	\\	
$					$		&$\vphantom{+}0	$	&s			&$	46(4)		$	&	$	4	$	\\	
$					$		&$\vphantom{+}0	$	&s			&$	53(4)		$	&	$	4	$	\\
$	\{{+}5/2,{-}5/2\}		$		&$\vphantom{+}0	$	&s			&$	31(4)		$	&	$	6	$	\\	
$	\{{+}7/2,{-}7/2\}		$		&$\vphantom{+}0	$	&s			&$	65.5(1.3)	$	&	$	11	$	\\	
$	\{{+}9/2,{-}9/2\}		$ 		&$\vphantom{+}0	$	&s			&$	35.9(0.4)	$	&	$	5.6	$	\\	
$					$		&$\vphantom{+}0	$	&s			&$	93.6(1.3)	$	&	$	17.3	$	\\	
$	\{{+}9/2,{-}7/2\}		$		&${+}1			$	&s			&$	66.3(0.7)	$	&	$	14.5	$	\\	
$	\{{+}7/2,{-}5/2\}		$		&${+}1			$	&s			&$	61(8)		$	&	$	29	$ 	\\	
$	\{{+}5/2,{-}3/2\}		$		&${+}1			$	&s			&$	23(8)		$	&	$	8	$ 	\\
$					$		&${+}1			$	&s			&$	53(8)		$	&	$	8	$ 	\\
$					$		&${+}1			$	&s			&$	122(8)	$	&	$	{>}40 $ 	\\		
$	\{{+}9/2,{-}5/2\}		$		&${+}2			$	&s			&$	159(3)	$	&	$	40	 $	\\	
$	\{{+}3/2,{+}3/2\}		$		&${+}3			$	&p			&$	76(4)		$	&	$	4	$	\\
$					$		&${+}3			$	&p			&$	140(4)	$	&	$	4	$	\\
$	\{{+}9/2,{-}1/2\}		$		&${+}4			$	&s			&$	61(8)		$	&	$	9	 $	\\	
$	\{{+}7/2,{+}7/2\}		$		&${+}7			$	&p			&$	105(3)	$	&	$	11	 $	\\	
$					$		&${+}7			$	&p			&$	182(2)	$	&	$	12	 $	\\	

\end{tabular}
}
\caption{
\textbf{Further measured loss resonances in \textsuperscript{40}K. }
Columns represent collision channel $\{m_1,m_2\}$ in the $f=9/2$ manifold,
quantum number $M= m_1+m_2$, the partial wave, the maximum loss
position $B_\text{loss}$, and the magnetic width $\sigma_\text{loss}$.
For these loss features no corresponding theoretical value exists.
Magnetic fields and widths are in Gauss.
The partial wave for each resonance has been assigned s-wave for equal losses in both components and p-wave if losses occur in only one component.
}
\label{losstab}
\end{table}

The collision channel $\{+1/2, -1/2\}$ is of particular interest. It is
the magnetic ground state of the spin subspace with zero magnetization $M=m_1+m_2=0$ and,
hence, losses due to inelastic two-body collisions can only occur by spin-relaxation from the weak magnetic dipole-dipole
interactions \cite{Ebling2014}.  
In this mixture, we have located a Feshbach resonance at $B_\text{res}^\text{calc}=389.6\,\text{G}$ with a
width of $\Delta B^\text{calc}=26.4 \, \text{G}$, which is about three
times larger than the width of the commonly used Feshbach resonances in
the channels $\{-9/2, -7/2\}$ and $\{-9/2, -5/2\}$.

From an experimental point of view, a broad resonance is desirable
as it lowers the technical demands for setting a stable value of
the interaction strength close to resonance.  Hence, the accurate
determination of the resonance position $B_\text{res}$ and the zero
crossing, where $a(B)$ has a node as a function of $B$, are of particular
importance. The latter occurs when $B_\text{zero}=B_\text{res}+\Delta B$
as can be seen from Eq.~\ref{eq:feshbachresonance} when $\gamma_2\to0$.
At this zero crossing atom loss tends to be small.  Recently, detection
methods such as radio-frequency spectroscopy \cite{Chin2004, Schunck2008},
collective excitations \cite{Bartenstein2004, Kinast2004}, Bloch
oscillations \cite{Gustavsson2008}, and spin segregation \cite{Du2008}
have been used to determine the resonance position or the zero crossing.
Here, we report on a method based on creating spin-wave excitations near a Feshbach resonance.
These excitations are sensitive to the sign of the scattering length $a(B)$
\cite{Miyake1985} and, in particular, the phase of the
spin-wave changes sign as the sign of $a(B)$ changes.  
Spin waves are an interaction-induced phenomenon and cannot be excited at the zero crossing.
For strong interactions close to the pole in $a(B)$ where the sign of $a(B)$ also changes many-body
effects induce additional corrections \cite{Trotzky2015}.  
Therefore, spin waves are particularly suited for finding the zero crossing of Feshbach resonances.

\begin{figure}[t]
	\centering
	\includegraphics[width=8.6cm]{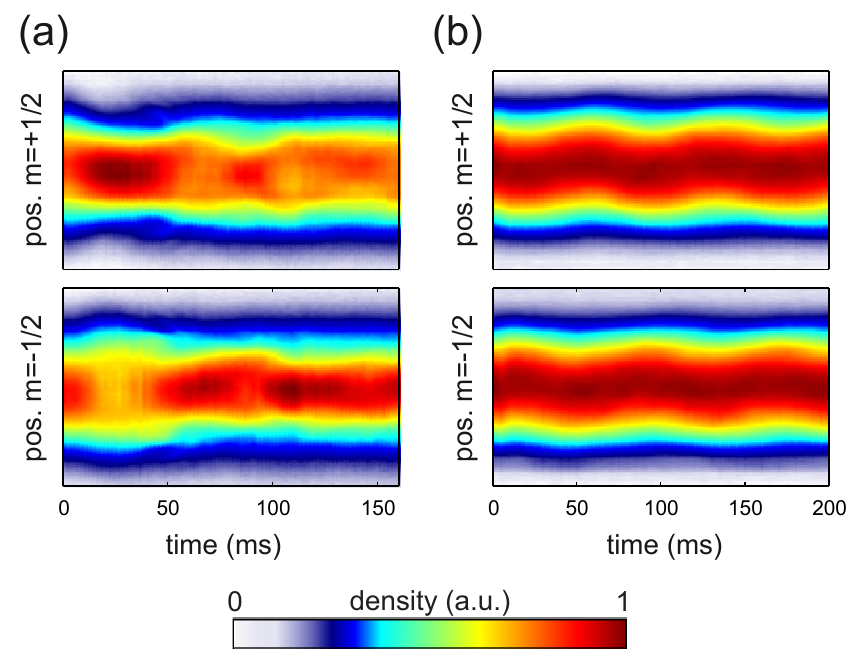}
	\caption{
	\textbf{Breathing and dipole mode resulting from a quadrupole
	and linear spin wave.} Time evolution of the density of the spin
	components $m=+1/2$ (top row) and $m=-1/2$ (bottom row), obtained
	by integrating over the two spatial directions orthogonal to the
	spin-wave excitation.  Panel a) shows quadrupole oscillations
	induced at $B=401 \, \text{G}$ for strong interactions and panel
	b) shows dipole oscillations induced at $B= 422\,\text{G}$ for weak
	interactions. Counterflow dynamics between the
	spin components can be observed in both panels.  }

	\label{fig:spinwaves_dipole_quadrupole}
\end{figure}

We excite spin waves with spatially-dependent magnetic fields that
induce spatially-dependent relative phase evolution between the two
spin components (for details see \cite{Heinze2013}).  
For this purpose, we first prepare a single-component Fermi gas in spin state $m=+1/2$
in an elongated dipole trap with trapping frequencies $\omega=2\pi
\times(70,70,12)\,\text{Hz}$ along the three independent spatial directions. 
At a magnetic field near the $\{+1/2,-1/2\}$ Feshbach resonance at
$B^{\rm calc}_{\rm res}=389.6$ G, we subsequently apply a radio-frequency pulse with a duration of $10\,\mu
\text{s}$ to create a coherent and equal superposition of the spin states
$m=+1/2$ and $-1/2$.  
We then excite a spin wave by using one of two types of field inhomogeneities along the weakest trapping direction. 
Close to the zero crossing, where the interaction
strength is small, we apply a linear magnetic gradient and excite linear
spin waves leading to dipole oscillations.
For larger interaction strengths small field inhomogeneities are sufficient.
Here, even the small residual magnetic quadrupole component originating from the Helmholtz coils excite quadrupole
spin waves and spatial breathing modes.
Examples of the spatial breathing and dipole modes are shown in
Figs.~\ref{fig:spinwaves_dipole_quadrupole}a) and b).
In both cases counterflow spin currents between the two spin components are induced,
eventhough the overall density remains constant.
While the dipole mode induced near the zero crossing of the resonance is long lived \cite{Heinze2013},
the breathing mode quickly decays due to incoherent collisions in the vicinity of $B_{\rm res}$ \cite{Koschorreck2013}.

\begin{figure}[t]
	\centering
	\includegraphics[width=8.6cm]{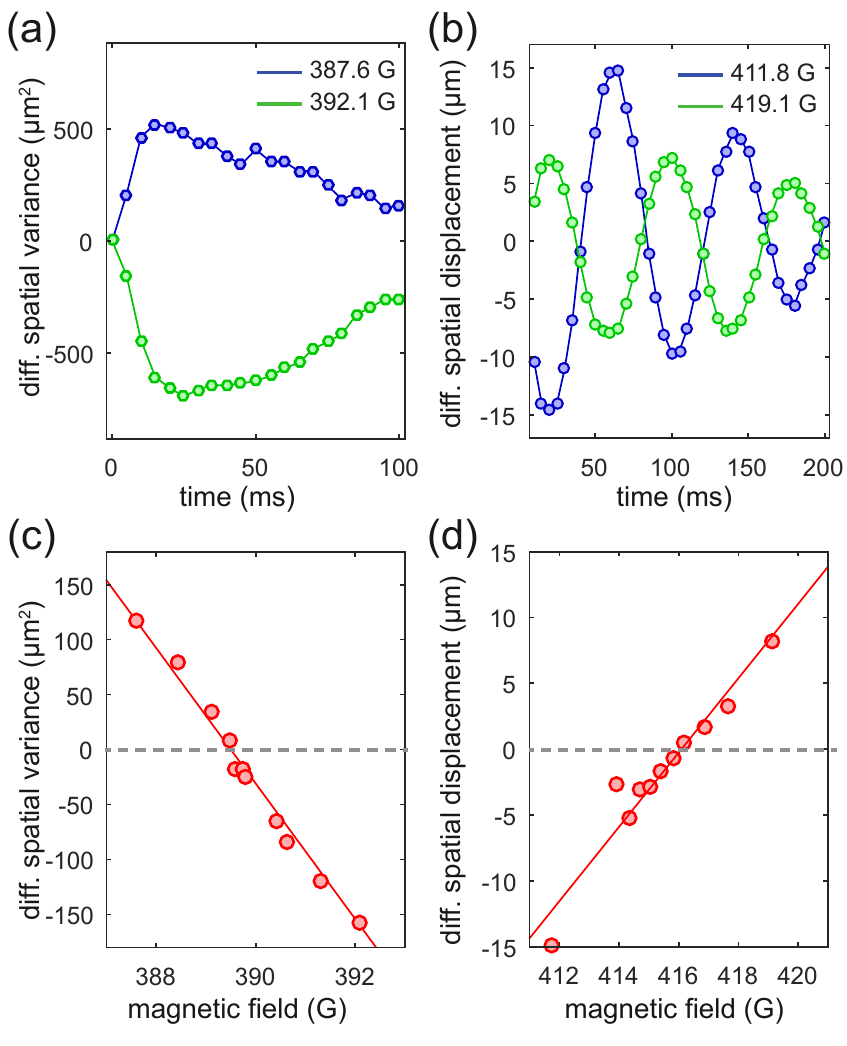}
	\caption{
	\textbf{Spin waves for different magnetic fields. }
(a) Time evolution of the differential variance for a breathing-mode spin
wave at two magnetic fields close the broad $\{+1/2,-1/2\}$ Feshbach resonance.  
 (b) Time evolution of the differential displacement for
a dipole-mode spin wave at two magnetic fields close to the zero crossing of this same
resonance.  
In (a) and (b) the solid lines are only guides to the eye. 
(c)  Maximum amplitude of the
differential variance as a function of magnetic field around
the resonance position.  
(d)  Amplitude of the differential displacement
 as a function of magnetic field near the zero crossing.
The amplitude is extracted by fitting a damped sine oscillation to data similar to \ref{fig:amplitude_quadrupole}(b).
The positions of the sign change of the scattering length are determined from a linear fit.
			}
	\label{fig:amplitude_quadrupole}
\end{figure}

The initial phase and amplitude of the breathing and dipole oscillation depends on the magnetic field strength.  
To extract this behavior from our data we analyze the time-dependence of the variance
of the spatial density profile along the weak trapping direction for the breathing mode and of the
displacement of each of the spin components for the dipole mode .
The time evolution of the difference in the variance of the two spin clouds is shown in Fig.~\ref{fig:amplitude_quadrupole}a)
for two magnetic fields on either side of the $B_{\rm res}=389.6$G resonance position.
Initially, the differential width grows up to a maximum value and then slowly decays to zero indicative of strongly damped motion.
Figure~\ref{fig:amplitude_quadrupole}b) depicts the time evolution of the difference in the displacement of the two spin components for
two magnetic fields on either side of the zero crossing.
The dipole oscillations remain visible over several periods.
The measurements in Figs.~\ref{fig:amplitude_quadrupole}a) and b) also reveal a strong magnetic field dependence.
Figures~\ref{fig:amplitude_quadrupole}c) and d) summarize this dependence as a function of magnetic field.
Using a linear fit, we can accurately determine the magnetic field at which the spin waves change their oscillation phase.
This yields 
$B_\text{sw}^\text{res}\,{=}\, 389.5\,(0.1)\,\text{G}$ and
$B_\text{sw}^\text{zero}\,{=}\,416.1\,(0.1)\,\text{G}$, respectively.
The quoted one standard deviation uncertainty follows from the fit.
We estimate a systematic error due to an uncertainty in the magnetic field calibration of $\Delta B_\text{sys} = 0.2 \,\text{G}$.

The measured value $B_\text{sw}^\text{zero}\,{=}\,416.1\,(0.1)\,\text{G}$
should coincide with the zero crossing of the scattering length.  In fact,
our measured value is in very good agreement with the theoretical
value of $416.0\,\text{G}$.  In contrast, the field value
$B_\text{sw}^\text{res}$ is affected by many-body effects and does
not serve as a precise measure for the Feshbach resonance position.
In future experiments, this could be overcome by using thermal gases,
where these effects are negligible \cite{Trotzky2015}.

In conclusion, we have observed twenty new Feshbach resonances in
\textsuperscript{40}K over a broad range of magnetic field values,
as well as twenty one further loss resonances whose origin has not yet
been theoretically determined.  Nineteen of the theoretically confirmed
resonances have $s$-wave character and one is a $p$-wave resonance.  Most
of these resonances are accompanied by losses.  In fact, these losses as
well as the elastic interactions can be tuned for each Feshbach resonance.
This allows for various future applications, such as the study of a quantum
Zeno insulator in optical lattices \cite{Syassen2008, Ripoll2009}.
Furthermore, a broad Feshbach resonance in the collision channel
$\{+1/2, -1/2\}$ has been identified at a magnetic field of
$B=389.6\,\text{G}$ with a width of $26.4\,\text{G}$, which
constitutes an ideal candidate for two-component studies  with
accurate control over the interaction strength.  In addition,
we demonstrated the creation of spin waves around this Feshbach
resonance, which allowed for a precise determination of the zero crossing.
Furthermore, we observed a phase shift of the spin waves near the Feshbach
resonance position, which might allow for the study of many-body effects
in strongly interacting Fermi gases in the future.

We acknowledge financial support by the Deutsche Forschungsgemeinschaft
(DFG) via Grant No.~FOR801 and DFG Excellence Cluster CUI: The Hamburg
Centre for Ultrafast Imaging, €"Structure, Dynamics, and Control of Matter
on the Atomic Scale. T.M.H. and L.C. acknowledge support from AFOSR MURI
Grant No. FA9550-09-1-0617.


\begin{thebibliography}{40}
\bibitem{footnote1}
Note that 17 of the 45 binary spin mixtures $\{m_1,m_2\}$ in the
$f=9/2$ manifold are stable against rapid losses due to inelastic two-body
spin-exchange collisions that conserve $M=m_1+m_2$ because of either the Pauli
exclusion principle or energy conservation.  This includes all binary
mixtures in spin states with $|m_1-m_2|=1$ or 2.  Out of these 17 spin
mixtures only $\{-9/2,-7/2\}$ is also stable against dipolar relaxation.
\end{thebibliography}
\end{document}